\documentstyle{mn}

\newif\ifAMStwofonts

\ifoldfss
  \ifCUPmtlplainloaded \else
    \NewTextAlphabet{textbfit} {cmbxti10} {}
    \NewTextAlphabet{textbfss} {cmssbx10} {}
    \NewMathAlphabet{mathbfit} {cmbxti10} {} 
    \NewMathAlphabet{mathbfss} {cmssbx10} {} 
  \fi
  \ifAMStwofonts
    \ifCUPmtlplainloaded \else
      \NewSymbolFont{upmath} {eurm10}
      \NewSymbolFont{AMSa} {msam10}
      \NewMathSymbol{\upi}     {0}{upmath}{19}
      \NewMathSymbol{\umu}     {0}{upmath}{16}
      \NewMathSymbol{\upartial}{0}{upmath}{40}
      \NewMathSymbol{\leqslant}{3}{AMSa}{36}
      \NewMathSymbol{\geqslant}{3}{AMSa}{3E}

       \let\le=\leqslant
      \let\geq=\geqslant 
    \fi
  \fi
\fi 

\ifnfssone
  \newmathalphabet{\mathit}
  \addtoversion{normal}{\mathit}{cmr}{m}{it}
  \addtoversion{bold}{\mathit}{cmr}{bx}{it}
  \newmathalphabet{\mathbfit} 
  \addtoversion{normal}{\mathbfit}{cmr}{bx}{it}
  \addtoversion{bold}{\mathbfit}{cmr}{bx}{it}
  \newmathalphabet{\mathbfss} 
  \addtoversion{normal}{\mathbfss}{cmss}{bx}{n}
  \addtoversion{bold}{\mathbfss}{cmss}{bx}{n}
  \ifAMStwofonts
    \ifCUPmtlplainloaded \else
      \UseAMStwoboldmath
      \makeatletter
      \new@mathgroup\upmath@group
      \define@mathgroup\mv@normal\upmath@group{eur}{m}{n}
      \define@mathgroup\mv@bold\upmath@group{eur}{b}{n}
      \edef\UPM{\hexnumber\upmath@group}
      \new@mathgroup\amsa@group
      \define@mathgroup\mv@normal\amsa@group{msa}{m}{n}
      \define@mathgroup\mv@bold\amsa@group{msa}{m}{n}
      \edef\AMSa{\hexnumber\amsa@group}
      \makeatother
      \mathchardef\upi="0\UPM19
      \mathchardef\umu="0\UPM16
      \mathchardef\upartial="0\UPM40
      \mathchardef\leqslant="3\AMSa36
      \mathchardef\geqslant="3\AMSa3E

       \let\le=\leqslant
      \let\geq=\geqslant 
    \fi
  \fi
\fi 

\ifnfsstwo
  \DeclareMathAlphabet{\mathbfit}{OT1}{cmr}{bx}{it}
  \SetMathAlphabet\mathbfit{bold}{OT1}{cmr}{bx}{it}
  \DeclareMathAlphabet{\mathbfss}{OT1}{cmss}{bx}{n}
  \SetMathAlphabet\mathbfss{bold}{OT1}{cmss}{bx}{n}
  \ifAMStwofonts
    \ifCUPmtlplainloaded \else
      \DeclareSymbolFont{UPM}{U}{eur}{m}{n}
      \SetSymbolFont{UPM}{bold}{U}{eur}{b}{n}
      \DeclareSymbolFont{AMSa}{U}{msa}{m}{n}
      \DeclareMathSymbol{\upi}{0}{UPM}{"19}
      \DeclareMathSymbol{\umu}{0}{UPM}{"16}
      \DeclareMathSymbol{\upartial}{0}{UPM}{"40}
      \DeclareMathSymbol{\leqslant}{3}{AMSa}{"36}
      \DeclareMathSymbol{\geqslant}{3}{AMSa}{"3E}

       \let\le=\leqslant
      \let\geq=\geqslant 
    \fi
  \fi
\fi 

\ifCUPmtlplainloaded \else
  \ifAMStwofonts \else 
    \def\upi{\pi}
    \def\umu{\mu}
    \def\upartial{\partial}
  \fi
\fi

\title{Modulation of AGN gamma rays by interaction with X-rays from an 
accretion disk hot spot}
\author[W. Bednarek and R.J. Protheroe]
       {W. Bednarek$^*$ and R.J. Protheroe \\
Department of Physics and Mathematical Physics,
The University of Adelaide, Adelaide, Australia 5005.\\
$^*$Permanent address: University of \L\'od\'z, 90-236\L\'od\'z, 
ul. Pomorska 149/153, Poland.  
               }
\date{University of Adelaide preprint ADP-AT-96-13, submitted to MNRAS}

\pubyear{199}

\begin{document}

\maketitle

\label{firstpage}

\begin{abstract}
We have developed a model for the variability of gamma ray
emission in jets of active galactic nuclei in which the
variability arises as a result of photon-photon pair production
interactions with X-rays emitted by a hot spot in the inner part
of the accretion disk.  As the hot spot orbits around the central
engine, the amount of absorption varies periodically. Our model
may account for the observed variability of TeV emission from
Markarian 421 and other blazars detected by the EGRET instrument
on the Compton Gamma Ray Observatory, as well as correlated
variations of X-ray and TeV gamma flux from Markarian 421.
Quasi-periodic variations observed in Markarian 421 enable us to
place an upper limit on the black hole mass of $(2-30)\times
10^8$ M$_\odot$.
\end{abstract}


\section{Introduction}

Two BL Lac objects, Markarian 421 and Markarian 501, have
recently been detected at TeV energies (Punch et al.~1992, Petry
et al.~1996, Quinn et al.~1996).  The observations of Markarian
421 show that the emission is variable on different time scales
from weeks to days (Kerrick et al.~1995, Macomb et al.~1996,
Schubnell et al.~1996, Buckley et al.~1996), and even variability
over a fraction of an hour has been observed (Gaidos et
al.~1996). In multiwavelength observations of Markarian 421,
simultaneous variations in X-rays and TeV $\gamma$-rays have been
reported (Takahashi et al.~1996a,b, Buckley et al.~1996). Also,
the X-ray flux shows quasi-periodic variability on a time scale
of $\sim$1 day superimposed on a $\sim 1$ week time scale decline
of the emission (Takahashi et al.~1996a).

This last feature is not easily explained in the models which
assume that X-rays and TeV $\gamma$-rays come from a
relativistic blob moving along the jet. However, it could be 
explained by a model in which the quasi-periodicity is caused by
the rotation of a massive star which enters the jet and creates a
shock at some distance from its surface, as recently proposed by
Bednarek \& Protheroe (1996).  Another natural explanation is
possible if the X-rays originate in a relatively small hot spot
rotating on the surface of the inner accretion disk. If the inner
disk thickness increases with radius, quasi-periodic modulation
of X-ray emission may be observed as the hot spot orbits 
around the black hole with the
disk, and is viewed at different inclination angles. The hot spot
emission may be also intrinsicly collimated by the hot spot's
geometry (e.g. the walls of the hot cavity in the disk), or by
an ordered magnetic field, and this may increase the level of
modulation. 

\begin{figure*}
\vspace{8cm}
\includegraphics{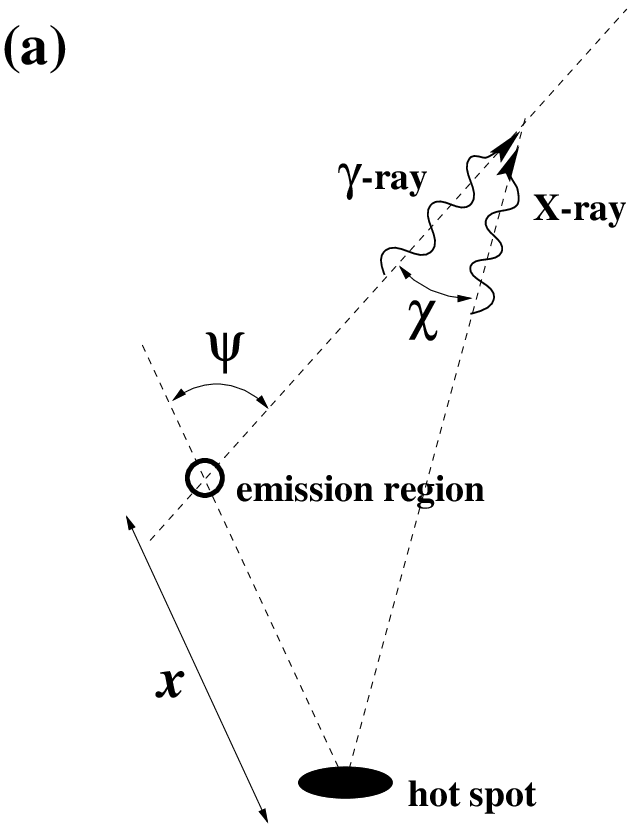}
\includegraphics{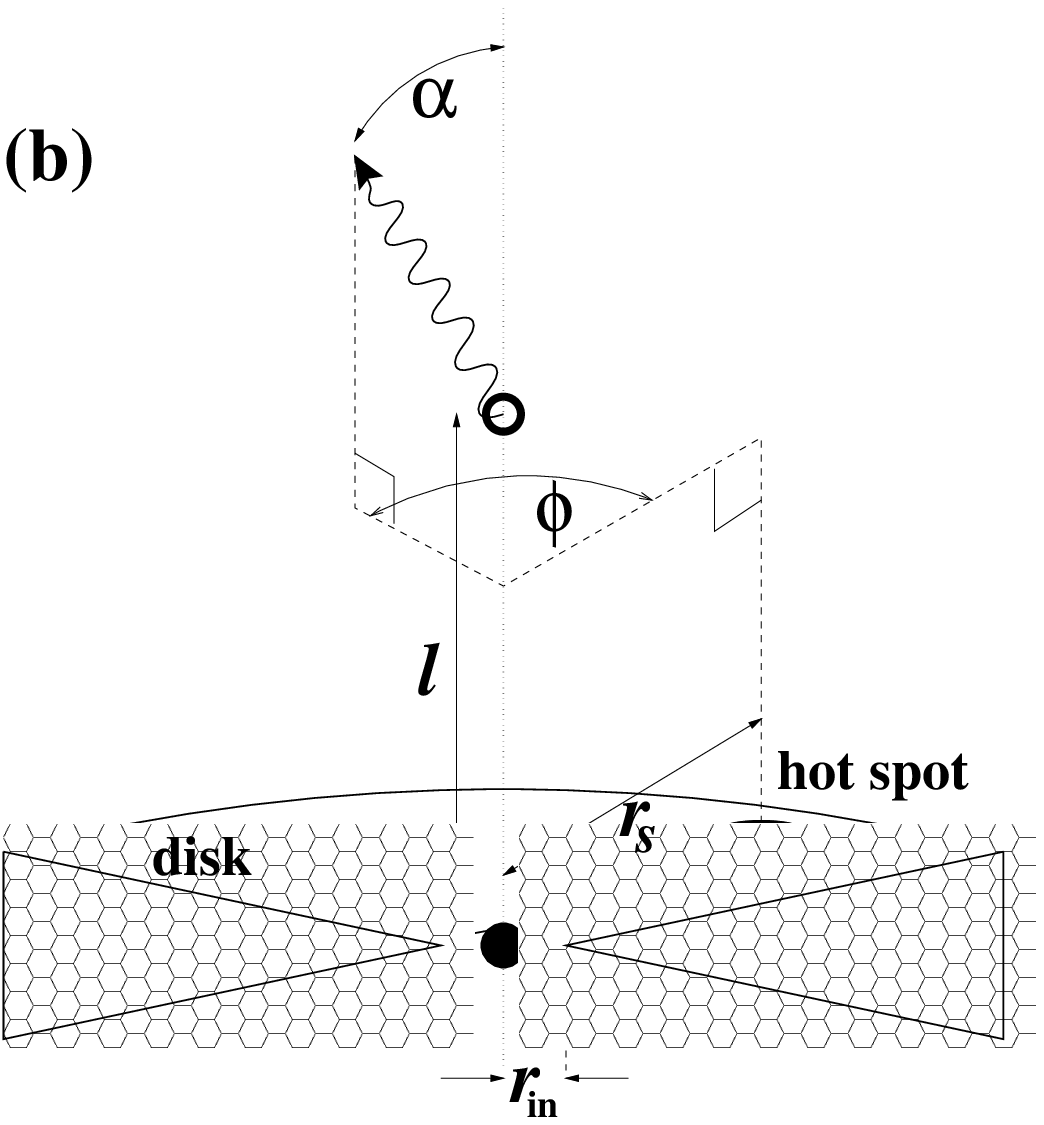}
\caption[]{(a) Simple picture of X-ray emitting hot spot and
emission region of VHE $\gamma$-rays located at a distance $x$
from the hot spot. The $\gamma$-rays, injected at an angle
$\psi$ to the direction from the hot spot, interact with the
X-ray radiation from the hot spot.  (b) Schematic representation
(not to scale) of the central region of AGN. The X-ray emitting
hot spot orbits with the accretion disk at a distance $r_s$ from
a massive black hole. The $\gamma$-rays emitted at angle $\alpha$
to the jet axis and at different distances $l$ along the jet can
interact with X-rays from the hot spot. The location of the hot
spot on the disk is defined by its phase $\phi$.}  \label{fig1}
\end{figure*}

Since the TeV emission can not originate in the hot spot itself 
(because of strong absorption), it must come from the jet
above the accretion disk. However the processes responsible for
the X-ray and TeV $\gamma$-ray variable emission can still be 
physically linked. It is very likely that significant changes 
in the jet conditions are initiated by changes in the accretion
conditions at the inner disk. For example, a sudden release of
energy in the inner disk region can change drastically the
viscosity parameter, propelling huge amounts of accreting matter,
and frozen-in magnetic field, into the black hole and/or the
jet. Such an energy release may be caused, for example by an explosive
reconnection of the magnetic field in the inner part of the
disk (e.g. Haswell, Tajima \& Sakai~1992), or as a result of 
a thermonuclear explosion on
the surface of a compact object (white dwarf, neutron star) which
has been captured by the accretion disk. If this happens, the
compact object can accumulate enough matter on its surface by
accretion in the dense medium to initialize a thermonuclear
outburst similar to those in scenarios proposed for nova
or Type I supernovae explosions (e.g. MacDonald 1983, Kovetz \& 
Prialnik~1994). 
The phenomena mentioned above can create a hot spot
(or a hot hole) on the surface of the disk. The energetic
particles in such a hot spot, bathed in the magnetic and
radiation field, may emit X-ray photons with the spectrum
observed during an outburst in Markarian 421.  The possible
occurrence of low photon energy outbursts in the inner accretion
disk has been recently proposed by B\"ottcher \&
Dermer~(1995) as a method for localizing the emission
region of VHE $\gamma$-rays in active galactic nuclei (AGN) or 
obtaining an upper limit
on the column density of matter around the disk if the low
energy photons are scattered by this matter.
    
The TeV emission could be produced in the
jet by any of the mechanisms suggested as being
responsible for $\gamma$-ray production in
blazars (e.g. Maraschi, Ghisellini \& Celotti~1992,
Mannheim \& Biermann~1992, Dermer \& Schlickeiser~1993,
Sikora, Begelman \& Rees~1994, Blandford \& 
Levinson~1995, Bednarek, Kirk \& Mastichiadis~1996,
Bloom \& Marscher~1996, Protheroe~1997).
If the TeV $\gamma$-rays are not produced too far away from the
hot spot, then they can be absorbed by photon-photon
pair production with X-ray photons from the hot spot, with the escape
probability depending on emission angle and energy. The
absorption strongly depends on the location of the TeV emission
region relative to the hot spot on the disk and the observer.
 
In this paper we discuss the effects of such possible selective
absorption on the escape of very high energy (VHE) $\gamma$-rays,
assuming a simple hot spot jet geometry. We start in Sect.~2 with
the simplest possible scenario: a relatively small X-ray emitting
hot spot and an emission region at some distance above the hot
spot which injects TeV $\gamma$-rays at some angle. We then
discuss in Sect.~3 a more complicated, but probably more
realistic, scenario in which the X-ray hot spot is created
somewhere in the inner disk and orbits with the disk around the
central black hole. Here, the TeV emission region extends along
the jet axis, and injects TeV $\gamma$-rays at some angle to the
jet axis. The $\gamma$ ray flux seen by a distant observer will
then be modulated with the orbital period of the hot spot by
selective absorption, unless the absorption of TeV $\gamma$-rays
in the accretion disk radiation dominates. We compute the optical
depth for TeV $\gamma$-rays in the accretion disk radiation, and
put the limits on the inner disk temperature, in Sect.~4. The
variability of TeV emission expected in such scenarios, and its
possible relevance to $\gamma$-ray modulation in compact sources,
e.g., VHE $\gamma$-ray blazars or galactic superluminal sources,
is discussed in Sect.~5.

\section{Optical depth of gamma-rays in the X-rays emitted by the 
hot spot} 

Let us assume that the X-rays are produced in a small hot spot,
and the TeV $\gamma$-rays are injected from the blob at a
distance $x$ above the hot spot, and propagate at an angle
$\psi$ measured from the direction defined by the hot spot and
the blob (see Fig.~1a). To obtain some idea about the possible
level of absorption of TeV $\gamma$-rays in a real source, we
assume that the hot spot emits X-rays with a broken power law
spectrum (phot cm$^{-2}$ s$^{-1}$ keV$^{-1}$), with an exponential 
cut-off below $\sim 20$ keV, similar to that observed during
an outburst in Markarian 421 (Takahashi et 
al.~1996a,b, Macomb et al.~1996, Buckley et al.~1996):
\begin{eqnarray}
F_X(\epsilon)  = \cases {0.187 \epsilon^{-1.9}, &  
                 $\epsilon<$  1.65  keV; \cr 
                 0.229 \epsilon^{-2.3}\exp{(\epsilon/20 \; {\rm keV})}, &  
		 $\epsilon\geq$ 1.65   keV. \cr }
\label{eq1}
\end{eqnarray} 
The optical depth for a $\gamma$-ray photon with energy $E_\gamma$,
measured from the point of injection to infinity, as a function
of the injection angle $\psi$, is given by
\begin{eqnarray}
\tau (E_\gamma, \psi) = {{d_{s}^2}\over{c x}} \int 
{{(1 - \cos\chi)}\over{D^2}} dH 
\int F_X(\epsilon)\sigma_{\gamma\gamma}(s) d\epsilon,
\label{eq2}
\end{eqnarray} 
\noindent
where $d_s\approx 180$ Mpc is the distance to Markarian 421, $D = (1 +
H^2 + 2H\cos\psi)^{-1/2}$, $\cos\chi =(H + \cos\psi)/D$, $H$
is the $\gamma$-ray propagation distance in units of $x$, $s = 2
\epsilon E_\gamma (1 - \cos\chi)$ is the center of momentum frame 
energy squared, $F_X(\epsilon)$ is given by
Eq.~(1), and $\sigma_{\gamma\gamma}(s)$ is the cross section for
photon-photon pair production (Jauch \& Rorlich~1980).

The results for three different $\gamma$-ray energies are shown
in Fig.~2. It is clear that, for reasonable hot spot-blob
distances, the interaction of TeV $\gamma$-rays with hot spot
X-rays can be important. For example, if $x = 10^{14}$ cm
$\gamma$-ray photons with energy $E_\gamma = 1$ TeV can escape
only within a cone with opening angle $\psi \sim
2^\circ$. For $x = 10^{17}$ cm, this escape cone has opening
angle $\sim 17^\circ$. Hence, if the X-rays are produced in a
relatively small region on the accretion disk, then the escape of
TeV $\gamma$-rays may be strongly dependent on the geometry of
the system. In the next section we discuss a possible geometry
for the emission regions of X-rays and VHE $\gamma$-rays which
may occur in the central regions of AGN.

\section{Hot spot orbiting around the central engine}

We now consider a more complicated scenario. Let us
assume that there is a small region (hot spot) in the inner disk
at a distance $r_s$ from the central black hole orbiting with 
angular velocity $\omega_s$ (see Fig.~1b). We neglect the hot
spot dimension which is assumed to be much smaller than the
characteristic distance scale, $r_s$. The angular velocity of the
hot spot is determined by $r_s$ and the black hole mass, and is
given by Kepler's law
\begin{eqnarray}
\omega_s  = (GM)^{1/2} r_s^{-3/2} = c (r_g/2)^{1/2} r_s^{-3/2},
\label{eq3}
\end{eqnarray}
\noindent
where $c$ is the velocity of light, and $r_g = 2GM/c^2$ is the 
Schwarzschild radius of a black hole with a mass $M$.
  
As mentioned above, we assume that the $\gamma$-rays are
continuously injected from the stream of plasma moving along the
jet, and $\alpha$ is the viewing angle with respect to the jet
axis (Fig.~1b). They may also originate in a blob, or a sequence
of blobs, moving along the jet. The TeV $\gamma$-rays move
through the X-ray radiation field of the hot spot located on the
disk surface at phase angle $\phi$ which depends on time,
\begin{eqnarray}
\phi = \phi_0 + \omega_s t,
\label{eq4}
\end{eqnarray}
\noindent
where $\phi_0$ is the initial phase of the hot spot (see
Fig.~1b), and we neglect any inward radial motion of the hot
spot.  As discussed below, this leads to a periodic variation of
the angle between the $\gamma$-rays propagating towards the
observer and the X-rays from the hot spot leading to periodic
variation of the $\gamma$-ray flux.  Kepler's law (Eq.~\ref{eq3})
gives us the radius of the hot spot in terms of the black hole
mass and the angular velocity,
\begin{eqnarray}
r_s  = (GM)^{1/3} \omega_s^{-2/3}, 
\label{eq3a}
\end{eqnarray}
which gives 
\begin{eqnarray}
r_s  = 1.4 \times 10^{14} M_8^{1/3} \;\;\; {\rm cm}, 
\label{eq3b}
\end{eqnarray}
for an orbital period of $\sim 1$ day, i.e. corresponding to the
observed quasi-periodic variability.  However, this hot spot
orbital radius must be at least that of the minimum inner
accretion disk radius, $\sim (0.5 - 3)r_g$, for maximally rotating Kerr and
Schwarzschild black holes respectively, giving
\begin{eqnarray}
r_s  > (1.5 - 9) \times 10^{13} M_8 \;\;\; {\rm cm} 
\label{eq3c}
\end{eqnarray}
Hence we obtain a strict upper limits to the mass of the black
hole in Markarian 421 of $M \le 2 \times 10^8$ M$_\odot$
(Schwarzschild black hole) or $M \le 3 \times 10^9$ M$_\odot$ (Kerr black
hole).
\begin{figure}
\vspace{7.2cm}
\includegraphics{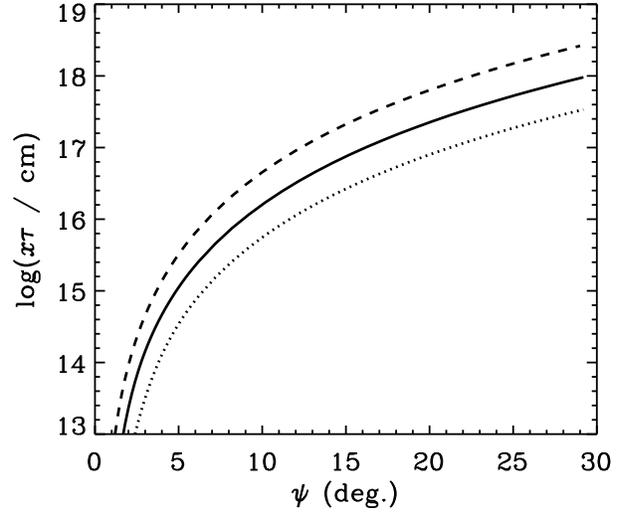}
\caption[]{The optical depth for the VHE $\gamma$-ray, 
multiplied by the distance $x$ of photon injection from the hot spot, 
versus the angle of photon 
injection, $\psi$. Separate curves correspond to different 
$\gamma$-ray energies: 0.32 TeV (dotted line), 1 TeV (full 
line), and 3.2 TeV (dashed line).}
\label{fig2}
\end{figure}
\begin{figure}
      \vspace{7.2cm}
\includegraphics{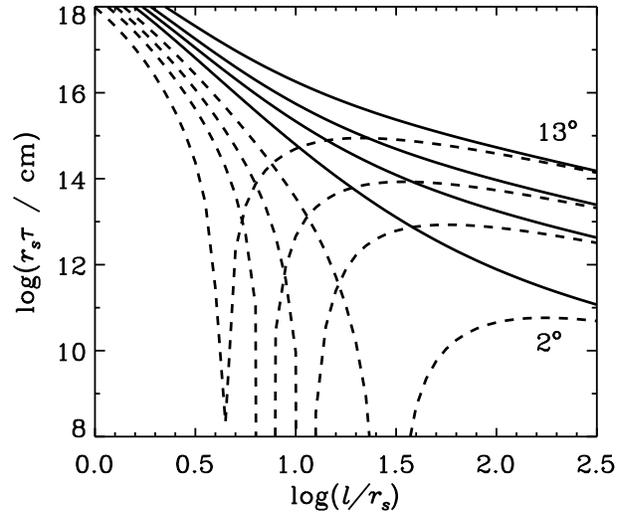}
      \caption[]{The optical depth, multiplied by the radius of the 
hot spot orbit $r_s$, for 1 TeV $\gamma$-rays 
as a function of distance $l$ of their injection point from the 
black hole, for angles of photon injection $\alpha = 2^\circ, 
5^\circ,8^\circ$, and $13^\circ$ degrees. 
Phase $\phi = 0^\circ$ is shown by the full lines and at 
$\phi = 180^\circ$ by the dashed lines.}
  \label{fig3}
\end{figure}

The optical depth for $\gamma$-ray photons can be calculated 
as a function of $r_s$, $\alpha$, $E_\gamma$, and $l$,
\begin{eqnarray}
\tau (E_\gamma, \alpha) = {{d_{s}^2}\over{c r_s}} \int 
{{(1 - \cos\chi)}\over{D^2}} dH 
\int F_X(\epsilon)\sigma_{\gamma\gamma}(s) d\epsilon,
\label{eq2mod}
\end{eqnarray} 
where it is convenient to express all distances 
in units of $r_s$: $H = h/r_s$, $L = l/r_s$. The values in 
Eq.~(\ref{eq2mod}) are given by,
\begin{eqnarray}
D = (1 + H^2 + L^2 + 2HL\sin\alpha \cos\phi)^{-1/2}, 
\end{eqnarray}
and 
\begin{eqnarray}
\cos\chi = (H + L\cos\alpha - \sin\alpha \cos\phi)/D. 
\end{eqnarray}

\begin{figure*}
\vspace{14.4cm}
\includegraphics{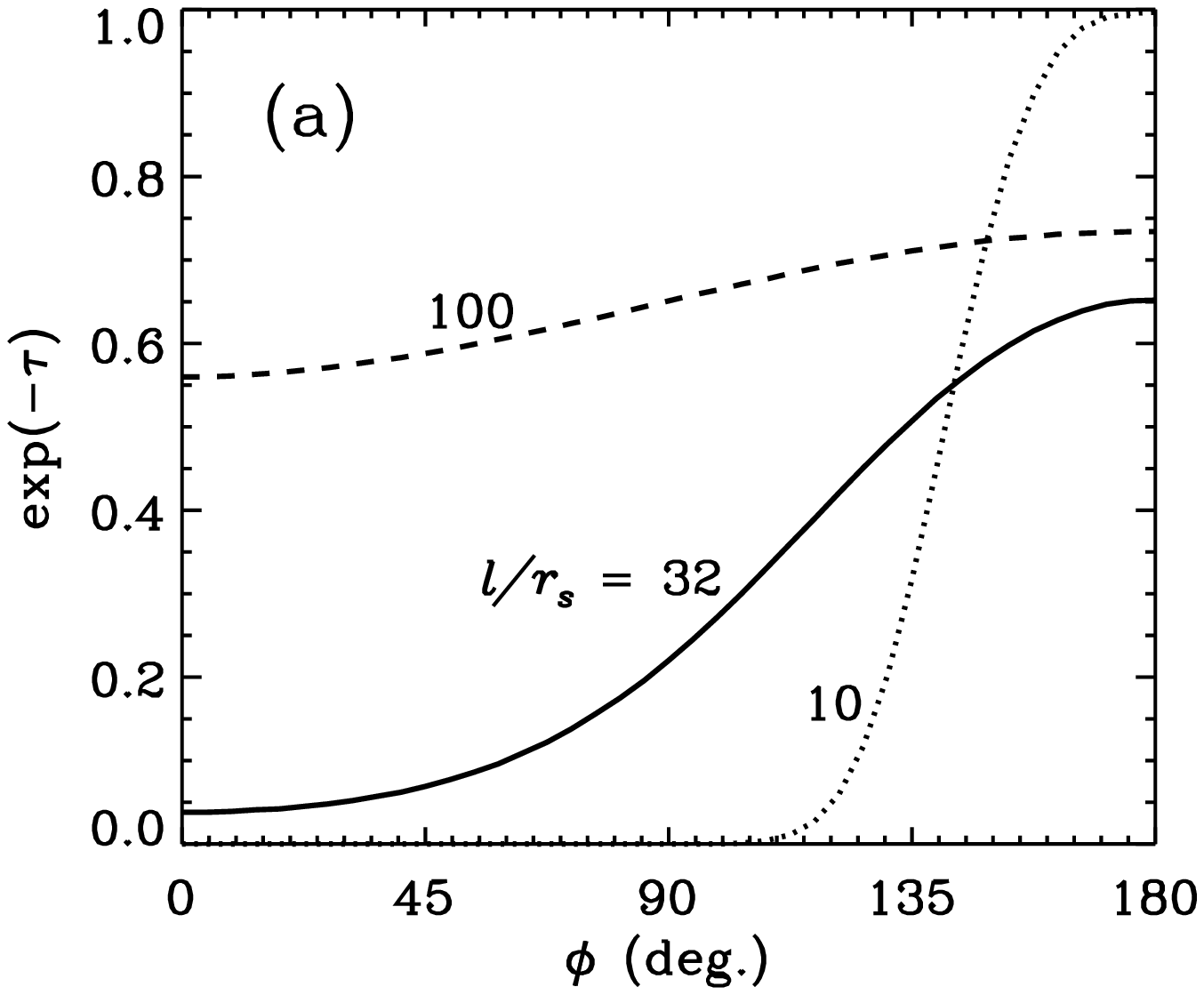}
\includegraphics{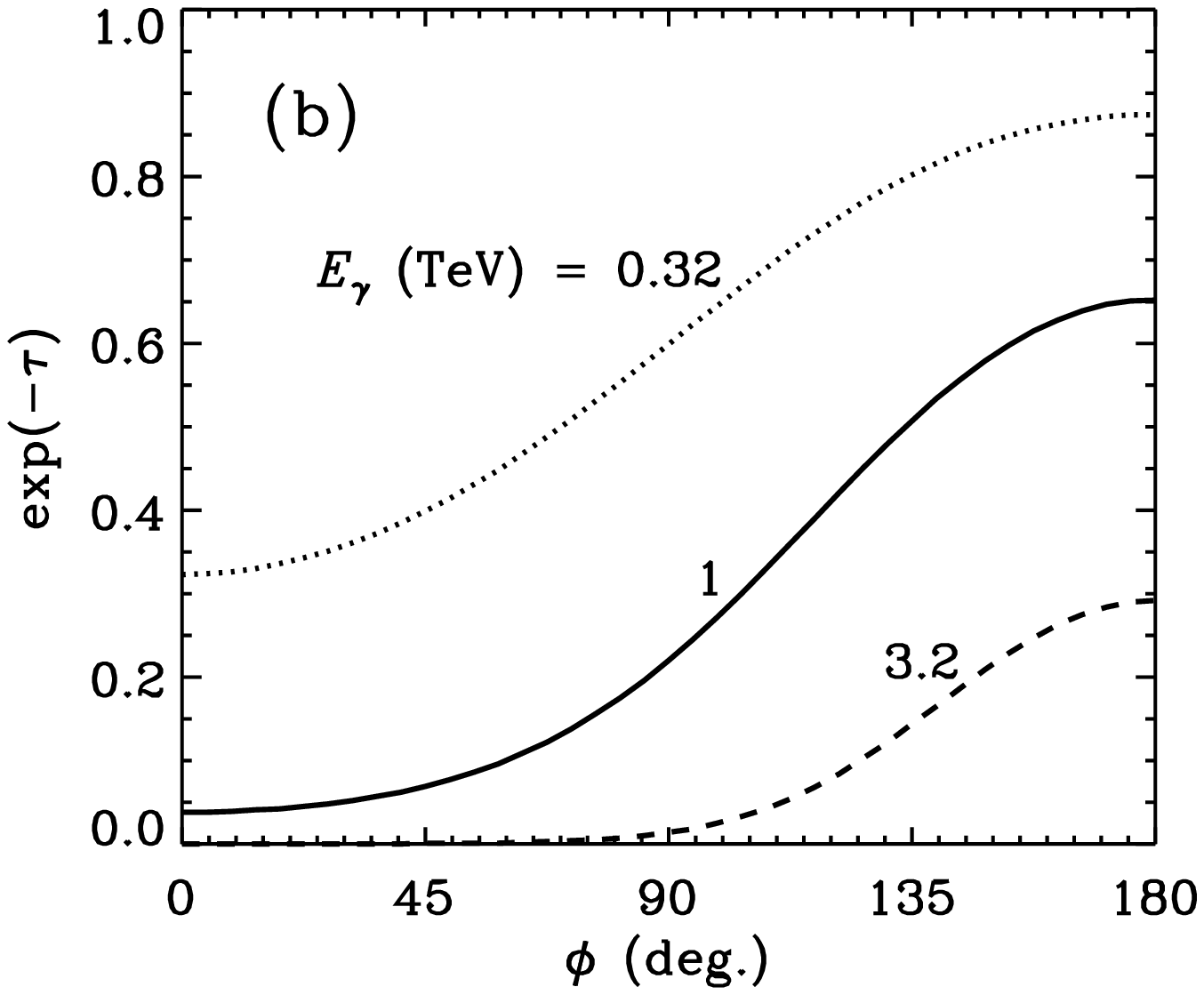}
\includegraphics{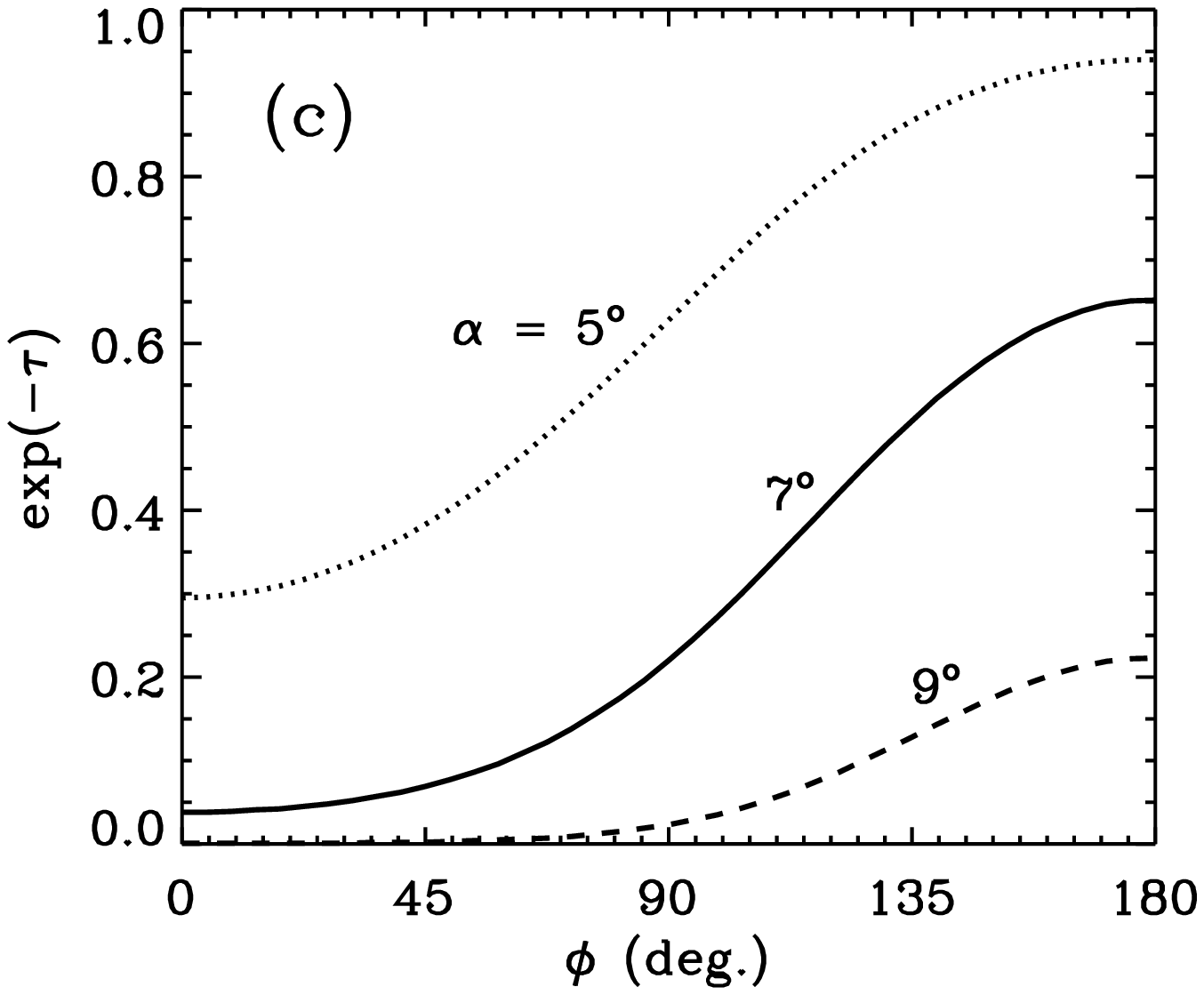}
\includegraphics{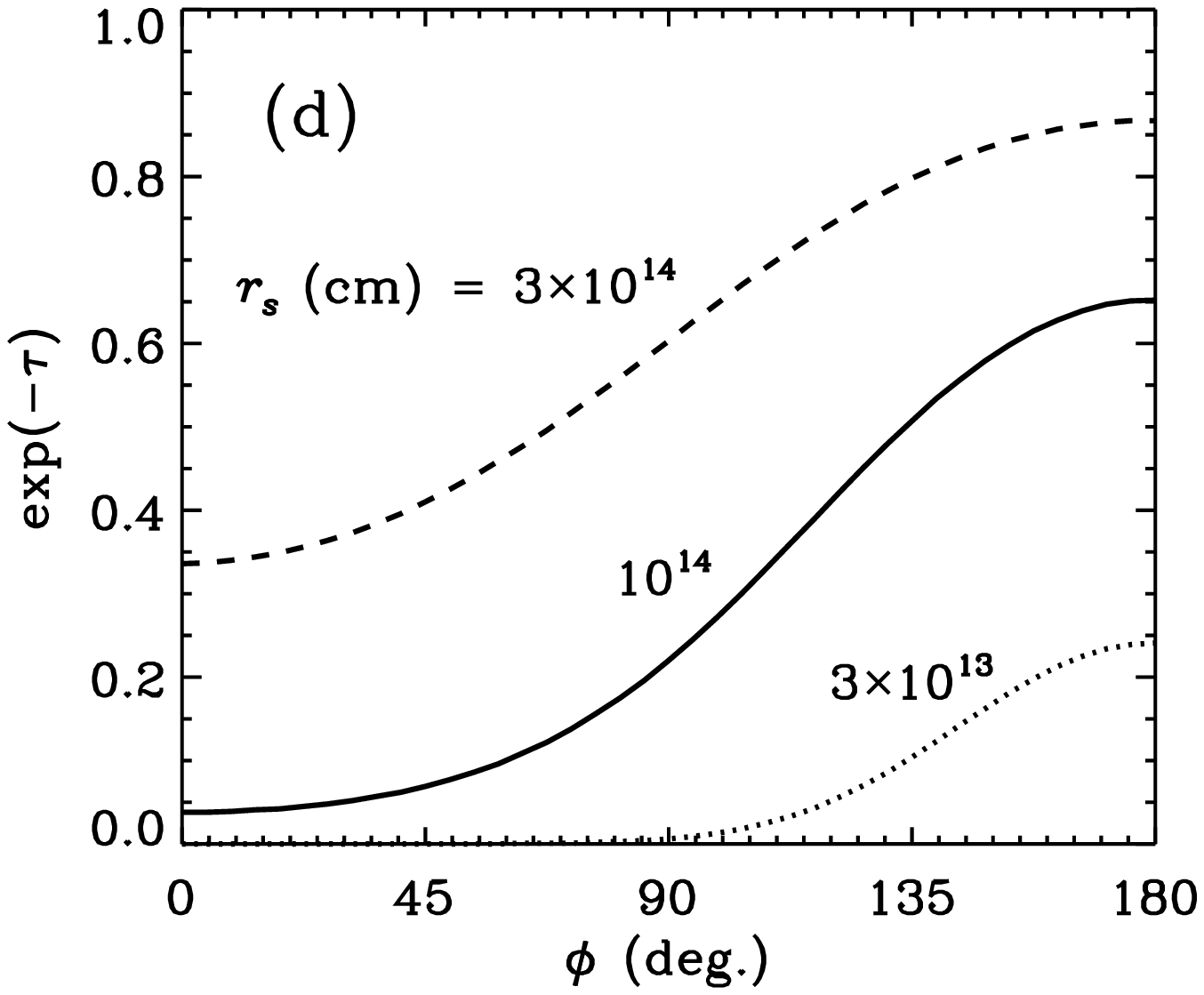}
\caption[]{Transmission probability as a 
function of phase $\phi$ of the hot spot. The $\gamma$-rays are 
injected at a distance $l$ and 
at angle $\alpha$. (a) Dependence of transmission probability on the 
hot spot phase for 1 TeV $\gamma$-rays, propagating at $\alpha = 
7^\circ$, for $r_s=10^{14}$ cm, and for log($l/r_s$) = 1 (dotted 
line), 1.5 (full line), and 2. (dashed line). (b) Dependence of 
transmission probability on 
$\gamma$-ray energy: $E_\gamma = 0.32$ TeV (dotted 
line), 1 TeV (full line), and 3.2 TeV (dashed line), with 
$r_s = 10^{14}$ cm, $\alpha = 7^\circ$, and log($l/r_s$) = 
1.5. (c) Dependence on the injection angle: $\alpha = 5^\circ$ 
(dotted line), $7^\circ$ (full line), and $9^\circ$ (dashed line), 
with $E_\gamma = 1$ TeV, $r_s = 10^{14}$ cm, and log($l/r_s$) = 1.5. 
(d) Dependence on the orbital radius of the hot spot: 
 $r_s = 3 \times 10^{13}$ cm (dotted 
line), $10^{14}$ cm (full line), $3 \times 10^{14}$ cm (dashed line), 
with $E_\gamma = 1$ TeV, $\alpha = 
7^\circ$, and log($l/r_s$) = 1.5}
  \label{fig4}
\end{figure*}

In order to gain an impression about the possible level of
modulation of the VHE $\gamma$-ray emission, we compute the
optical depth at $E_\gamma = 1$ TeV for two critical locations of
the hot spot in its orbit around the black hole: $\phi =
0^\circ$, and $180^\circ$. The results are shown in Fig.~3 as a
function of the distance of the $\gamma$-ray injection point from
the black hole $l$, for four different angles of observation
$\alpha$.  If the hot spot is located at phase $\phi = 0^\circ$,
and $\alpha$ is large, then the optical depth drops off
monotonically with distance. The dependence of the optical depth
on $l$, for when the hot spot is located at $\phi = 180^\circ$,
shows characteristic cusps at distances which correspond to the
$\gamma$-ray photons and the X-ray photons from the hot spot
moving such that the angle between their directions is very
small. The location of the cusps shifts to smaller $l$ with
increasing $\alpha$. The optical depths for $\phi=0^\circ$ and
$\phi=180^\circ$ converge at large distances $l$ from the black
hole because the angles between interacting photons become
comparable. The largest modulation of the $\gamma$-ray signal is
expected at distances for which the differences between the full
and dashed lines, computed for the same value of $\alpha$, are
highest, and have $\tau \gg 1$ for $\phi=0^\circ$ (full lines)
and $\tau \ll 1$ for $\phi=180^\circ$ (dashed lines).

Closer inspection of Fig.~3 allows us to choose the parameters in
this scenario ($r_s$, $l$, $\alpha$) for which absorption effects
can significantly modulate the 1 TeV $\gamma$-ray signal with the
orbital period of the hot spot.  In order to show the level of
possible modulation, we compute the transmission probability for
$\gamma$-ray photons ($e^{-\tau}$), selecting as a basic example
the case where $E_\gamma = 1$ TeV, log($l/r_s$) = 1.5, and $r_s =
10^{14}$ cm.  We investigate the dependence of the transmission
probability as a function of phase $\phi$ for different values of
the parameters mentioned above. Fig.~4a shows the level of
modulation for 1 TeV $\gamma$-rays injected at angle $\alpha =
7^\circ$, for three distances from the black hole: log($l/r_s$) =
1.0, 1.5, and 2.0. At $l = 10r_s$ (dotted line in Fig.~4a), the
modulation is very strong and, in fact, the 1 TeV $\gamma$-rays
can only escape when the hot spot is close to phase $\phi =
180^\circ$. At larger distances the level of modulation is lower
(a factor of $\sim 15$ for $l\approx 30r_s$; full line in
Fig.~4a), and becomes negligible at $l = 100r_s$ (dashed line).

The dependence of the transmission probability on the
$\gamma$-ray energy is shown in Fig.~4b. The transmission
probability for low energy $\gamma$-rays is high, but the level
of modulation of the $\gamma$-ray signal by the orbital period of
the hot spot is low. For example, for $E_\gamma = 0.32$ TeV, the
modulation is only by a factor of $\sim 3$, in comparison to a
factor of $\sim 15$ at 1 TeV and much higher at 3.2 TeV. The
transmission probability depends strongly on the angle of
$\gamma$-ray injection (see curves for $\alpha = 5^\circ,
7^\circ$, and $9^\circ$ in Fig.~4c).  However, the general
pattern of modulation of the $\gamma$-ray signal with orbital
phase is conserved. In Fig.~4d we show the dependence of the
transmission probability on the orbital radius of the hot spot
for $E_\gamma = 1$ TeV, $\alpha = 7^\circ$, and log($l/r_s$) =
1.5. As expected, when the transmission probability is lowest
(for a more compact source; see dotted line for $r_s = 3\times
10^{13}$ cm) the level of modulation of the $\gamma$-ray
signal with the orbital period is highest.

One must keep in mind that, in the present scenario, the variable
signal will be multiplied by the $\gamma$-ray emission efficiency
which is likely to decline with the distance along the jet
axis. However, the effects described above may be irrelevant if
the radiation from the accretion disk in which the hot spot is
located, is too intense. In the next section we shall discuss
possible absorption of the TeV $\gamma$-ray signal in the
accretion disk radiation.

\section{Absorption of VHE gamma rays in the disk radiation}

Within the central regions of the AGN, interactions with photons
directly from the accretion disk can be important (Becker and
Kafatos 1995, Bednarek 1996; see Bednarek 1993 for an earlier
discussion of $\gamma$-ray escape from the radiation field of an
accretion disk surrounding a neutron star in an X--ray binary
source).  In order to check for which accretion disk parameters
the modulation effects of VHE $\gamma$-ray emission by the X-rays
from the hot spot, discussed in Sect.~3, are not swamped by
absorption in the accretion disk radiation, we compute the
optical depth, $\tau$, for $\gamma$-rays in the disk radiation
(e.g. Bednarek~1993, Protheroe \& Biermann~1996). We assume the
surface of the accretion disk emits black body radiation with
temperature varying with radius as $T(r) = T_{in}
(r/r_{in})^{-3/4}$, where $T_{in}$ is the temperature at the
inner disk radius $r_{in}$ (Shakura \& Sunyaev~1973). The VHE
$\gamma$-rays can escape if they are injected at distances
farther than the so-called the ``radius of the $\gamma$-sphere'',
$r_\gamma$, which is defined by the following condition
(e.g. Blandford \& Levinson~1995, Bednarek~1996),
\begin{eqnarray}
\tau(E_\gamma, \alpha, r_\gamma) = 1,
\label{eq}
\end{eqnarray}
\noindent
where $\tau$ is the optical depth to photon-photon pair
production for $\gamma$-rays propagating at angle $\alpha$,
measured from the disk axis, through the radiation field of the
accretion disk.

The radius of the $\gamma$-ray photosphere, expressed in units of
$r_{in}$, is shown in Fig.~5 as a function of $\gamma$-ray energy
for different inner disk temperatures and radii.  The angle of
$\gamma$-ray injection has been fixed at $\alpha = 7^\circ$
because our results for the modulation of the $\gamma$-ray signal
presented in Figs.~4(a)-(d) have been shown mainly for this
angle. As expected, the radius of $\gamma$-ray photosphere
increases with $\gamma$-ray energy, and also with intensity of
the radiation field of the disk (i.e. higher $T_{in}$ and/or
$r_{in}$).  For the modulation of VHE $\gamma$-rays by
interaction with X-rays from an accretion disk hot spot to take
place the emission point of VHE $\gamma$-rays must be outside the
$\gamma$-ray photosphere, i.e $l > r_\gamma(E_\gamma)$.

\section{Conclusion}

We have shown that if the X-ray emission observed from Markarian
421 during the outburst stage (Takahashi et al.~1996a,b)
originates on the surface of the accretion disk, then it creates
a sufficiently strong radiation field for the absorption of VHE
$\gamma$-rays provided they originate in the jet near the
accretion disk (Sect.~2).  We have discussed a simple geometrical
model in which a small hot spot emitting X-rays orbits with the
accretion disk around the central black hole, and the VHE
$\gamma$-rays are injected at some distance along the jet at a
small angle to its axis. We have shown that absorption effects on
VHE $\gamma$-rays in the radiation field of a hot spot, with an
X-ray spectrum similar to that of Markarian 421, can become very
important for some sets of model parameters, i.e. angle of
injection $\alpha$, location of injection point $l$, and radius
of the orbit of the hot spot $r_s$ (see Fig.~3). Absorption
becomes stronger for higher energy $\gamma$-rays and may
result in a change in the shape of the VHE $\gamma$-ray spectrum.

Modulation of the $\gamma$-ray emission with the orbital period
of the hot spot is also very significant, and may result in
simultaneous quasi-periodic oscillations of X-ray and TeV
$\gamma$-ray emission. Note that $\sim$ 1 day quasi-periodic
variability has been detected by ASCA in X-rays from Markarian
421 (Takahashi et al.~1996a), and enables us to place an upper
limit on the black hole mass for the present scenario of $M \le 2
\times 10^8$ M$_\odot$ (Schwarzschild black hole) or $M \le 3
\times 10^9$ M$_\odot$ (Kerr black hole). Modulation effects such
as those described in the present work are expected provided the
disk radiation itself does not produce a sufficiently intense
radiation field for VHE $\gamma$-ray absorption. For the
parameters discussed above, the disk radiation is not important
if the disk inner temperature is less than $\sim 2\times 10^4$ K
for a disk inner radius $r_{in} = 10^{14}$ cm.  The direct
accretion disk radiation is observed in some quasars e.g., in
quasar 3C 273, which is a $\gamma$-ray emitting active galaxy and
has an inner disk temperature $(2 - 3) \times 10^4$K
(Shields~1978, Malkan \& Sargent~1982).  The disk radiation
fields of the BL Lacertae objects, e.g. Markarian 421, are
significantly weaker since their spectra are continuous, without
strong evidence of emission lines or other thermal features.

\begin{figure}
\vspace{7.2cm}
\includegraphics{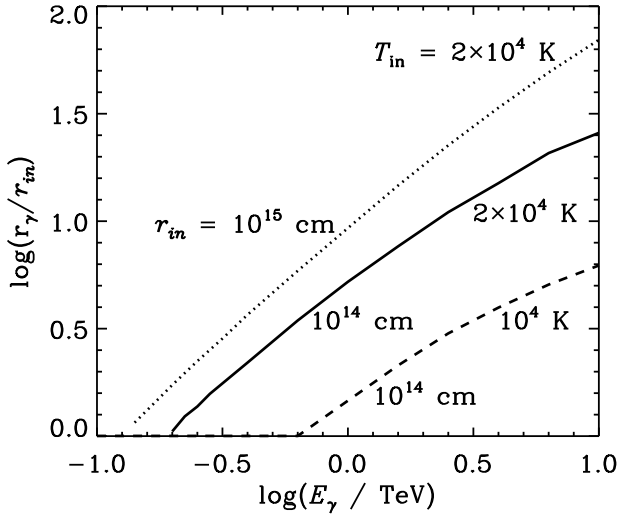}
       \caption[]{
Radius of the $\gamma$-ray photosphere $r_\gamma$, expressed in 
units of the disk inner radius $r_{in}$, as a function of the energy 
of $\gamma$-ray photons. The disk is a Shakura-Sunyaev type
(Shakura \& Sunyaev~1973) 
with the temperature $T_{in}$ at the inner radius $r_{in}$. Separate 
curves correspond to different temperatures at the disk inner radius: 
$T_{in} = 2\times 10^4$ K, and $r_{in} = 10^{14}$ cm 
(full curve); $T_{in} = 2\times 10^4$ K, and $r_{in} = 10^{15}$ cm 
(dotted curve); $T_{in} = 10^4$ K, and $r_{in} = 10^{14}$ cm (dashed 
curve).}
\label{fig5}
    \end{figure}

The TeV $\gamma$-rays may also be absorbed on radiation scattered
by matter distributed around the accretion disk (e.g. Sikora,
Begelman \& Rees~1994, B\"otcher \& Dermer~1995). However, these
radiation fields, if important, are difficult to consider in
detail because they depend on parameters which are not known
precisely.  In some sources, where the bulk of the infrared
emission comes from a dusty molecular torus surrounding the
central engine, VHE emission from points below the top of the
torus would be absorbed by photon-photon pair production on
infrared radiation from the torus (Protheroe and Biermann 1996).
However, in the absence of such a torus, or if the temperature of
the torus is low, modulation of VHE $\gamma$-rays by interaction
with X-rays from an accretion disk hot spot as described in the
present paper should be possible.

The stream of relativistic jet plasma, or sequence of blobs, from
which the VHE $\gamma$-rays might originate does not need to move
rectilinearly along the jet axis. It may emerge from the region
of the hot spot and move outwards from the disk following, e.g.,
helical magnetic field lines. Farther from the disk, the angular
rotation of the plasma stream can be slower than that of the
inner disk. In such a situation the period of the hot spot's
orbit, responsible for the modulation of observed X-ray flux in
this scenario, may be shorter than the period of the TeV
$\gamma$-ray modulation.

An interesting feature of the present scenario is that it
predicts a time delay of the peak TeV $\gamma$-ray flux with
respect to the peak X-ray flux.  This is caused by
the different paths to the observer from the X-ray and
$\gamma$-ray emission regions. The delay depends on the
distance of the $\gamma$-ray emitting region from the disk $l$,
the radius of the hot spot's orbit $r_s$, and the angle
of observation $\alpha$, and is given by
\begin{eqnarray}
\Delta t = (r_s^2 + l^2)^{1/2} [1 - \cos(\theta - \alpha)]/c,
\label{eq6}
\end{eqnarray} 
\noindent
where, $\tan \theta = r_s/l$.  For example, if $r_s = 10^{14}$
cm, $\alpha = 7^\circ$ and $l = 30r_s$, then the expected delay
of the maximum in $\gamma$-ray flux is $\Delta t\approx 40$
minutes, and for $l = 10r_s$ the delay is $\Delta t \approx 6$
minutes.
\section*{Acknowledgements}

W.B. thanks the Department of Physics and Mathematical Physics at
the University of Adelaide for hospitality during his visit. We
thank Qinghuan Luo for reading the manuscript.  This research is
supported by a grant from the Australian Research Council.

\end{document}